# Manipulating matter with strong coupling: harvesting triplet excitons in organic exciton microcavities

Daniel Polak[1], Rahul Jayaprakash[1], Anastasia Leventis[2], Kealan J Fallon[2], Harriet Coulthard[1], Anthony J Petty II[3], John Anthony[3], Hugo Bronstein[2], David G Lidzey[1], Jenny Clark[1]* and Andrew J Musser[1]*

[1] Department of Physics and Astronomy, University of Sheffield, Hicks Building, Hounsfield Road, Sheffield S3 7RH, UK

[2] Department of Chemistry, University of Cambridge, Lensfield Road, Cambridge CB2 1EW, UK

[3] Department of Chemistry, University of Kentucky, Lexington, Kentucky 40506-0055, USA

*email: a.musser@sheffield.ac.uk, jenny.clark@sheffield.ac.uk

**Abstract**

Exciton-polaritons are quasiparticles with mixed photon and exciton character that demonstrate rich quantum phenomena, novel optoelectronic devices and the potential to modify chemical properties of materials. Organic semiconductors are of current interest for their room-temperature polariton formation. However, within organic optoelectronic devices, it is often the 'dark' spin-1 triplet excitons that dominate operation. These triplets have been largely ignored in treatments of polariton physics. Here we demonstrate polariton population from the triplet manifold via triplet-triplet annihilation, leading to polariton emission that is longer-lived (>µs) even than exciton emission in bare films. This enhancement arises from spin-2 triplet-pair states, formed by singlet fission or triplet-triplet annihilation, feeding the polariton. This is possible due to state mixing, which —in the strong coupling regime— leads to sharing of photonic character with states that are formally non-emissive. Such 'photonic sharing' offers the enticing possibility of harvesting or manipulating even states that are formally dark.

**Introduction**

The exploration of new material properties typically faces significant practical constraints from cumbersome synthesis and morphological control. In recent years, however, it has been shown that many materials properties can be non-synthetically tuned with confined light fields to form exciton-polaritons[1–3], pointing the way to an entirely new field of microcavity-controlled materials[4–8]. These exciton-polaritons are quasi-particles mixing light (photon) and matter (exciton) components, leading to rich quantum effects[9–13] and potential optoelectronic applications[2–4,14–18]. Exciton-polaritons are formed by placing a semiconductor between two metal mirrors to create a Fabry-Perot microcavity in which light of the correct angle and wavelength can be trapped (Fig 1a). If the material within the cavity has a strong exciton absorption, in resonance with the trapped photon mode, the exciton and photon can couple and form hybrid polariton states (Fig 1b). As a consequence of the mixed exciton-photonic character of these states, a measurement of reflected light as a function of incident angle demonstrates the typical dispersion shown in Fig 1c, with the upper polariton branch (UPB) and lower polariton branch (LPB) split around the excitonic energy.

Most studies of exciton-polariton physics have focussed on inorganic semiconductor systems[9,10,13–15]. In comparison, organic semiconductors have the advantage of high oscillator strength[19], which leads to Rabi splittings in the range 0.1-1eV[1,2,19–22]. Organic semiconductors also have low dielectric constants ($\varepsilon_r$ typically 2-4). Consequently, photoexcitation results in bound electron-hole pairs known as Frenkel excitons, with binding energies on the order of 0.5-1eV. Such high binding energies allow for room temperature polariton formation and condensation, the latter observed now in several organic semiconductor microcavities[11,23,24]. The tightly bound Frenkel excitons also exhibit complex photophysics, with numerous radiative and non-radiative decay pathways possible following initial photoexcitation (Fig 1d)[25]. These pathways are rarely treated in detail in organic exciton-polariton studies, where the focus is primarily on 'bright' singlet (spin-0) excitons. However, intermolecular relaxation to form weakly emissive excimers can significantly influence microcavity emission dynamics[26], and theoretical attention increasingly has started to focus on the impact of other non-radiative photophysical processes[6–8].

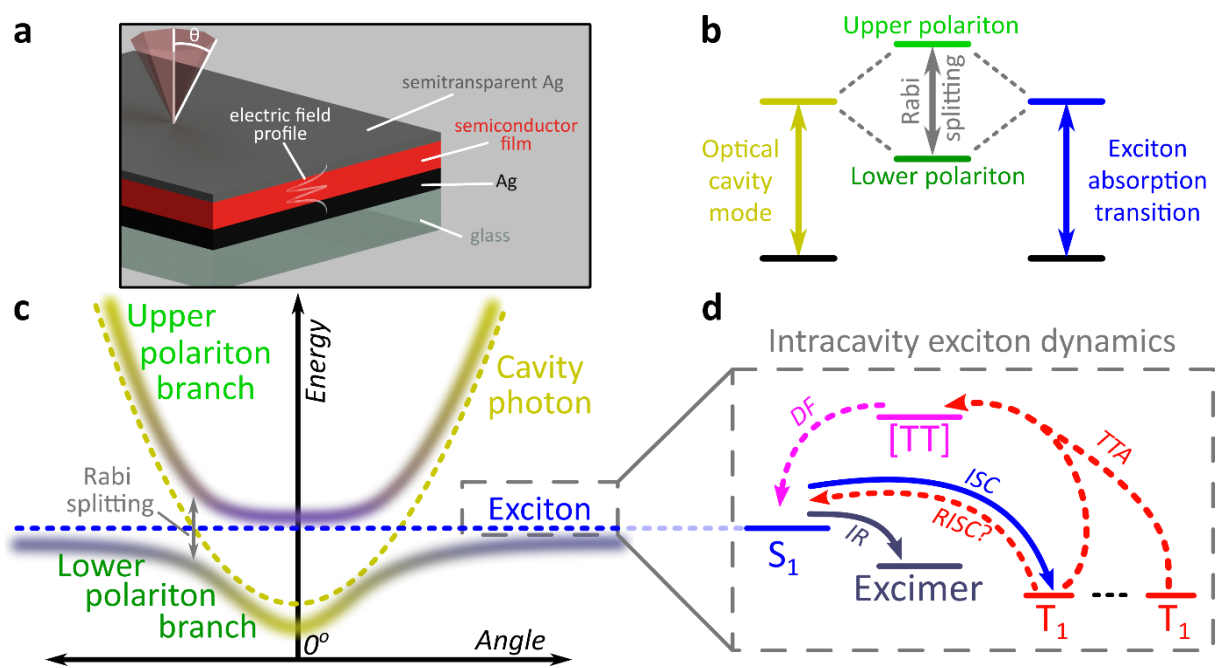

**Fig 1| Strong light-matter coupling in optical microcavities. a** Microcavity structure. A thin film of organic semiconductor or dye dispersed in neutral polymer matrix is deposited between two mirrors, here Ag. The thickness determines the energy of the confined photonic mode and thus the profile of the electric field inside the cavity, shown here for the λ-mode. Reflection and emission from the cavity are measured as a function of angle θ, with 0° defined as normal to the cavity surface. **b** When the cavity mode and the excitonic transition of the semiconductor are near resonance, these two states can couple, forming hybrid upper and lower polariton states. **c** Unlike the exciton (blue), the cavity mode (gold) exhibits distinct angular dispersion. Coupling between the two yields dispersed polariton branches, with characteristic anti-crossing at the exciton energy. Shading indicates the degree of photonic (gold) vs excitonic (blue) character in the state. **d** Typical excitonic processes possible within organic semiconductor films. IR: intermolecular relaxation, (R)ISC: (reverse) intersystem crossing, TTA: triplet-triplet annihilation, DF: delayed fluorescence. Solid arrows indicate processes known to modify exciton-polariton emission dynamics, while dashed arrows show processes not explored within microcavities.

We focus here on the role in these systems of triplet (spin-1) excitons. An additional consequence of the low dielectric constant of organic materials is a large exchange energy, which results in the lowest triplet exciton being located >0.5eV below the first singlet exciton[27]. Triplet excitons and their management are critical in organic semiconductor devices such as displays and solar cells[27–30]. For example, 75% of excitons formed by electron-hole recombination in optoelectronic devices are triplets due to spin statistics. Triplets are a main reason for the absence of a continuous optically pumped or electrical-injection lasers, but could be useful in solar cells[29,30]. Triplets can be generated from singlet excitons via intersystem crossing, which is generally slow (ns or longer) due to weak spin-orbit coupling. Once formed, return to the ground-state requires a spin-flip. Therefore, triplets are non-emissive and long-lived (>>µs).

It is generally thought that only states with large oscillator strength couple to the photon in a microcavity, with triplet states considered a loss channel in organic exciton-polariton systems[22]. Because of progress in electrically injected polariton devices[15–18], however, it is important to consider in more detail the fate of these states and how they interact with polaritons. Similarly to electrical injection, a very large reservoir of triplets can be generated by photoexcitation in some materials. Large triplet populations can be optically generated by using systems with strong spin-orbit coupling resulting in fast intersystem crossing[22] or systems in which the exchange energy is so large that the singlet energy is approximately twice the triplet energy. In the latter, photoexcitation into the singlet state results in formation of two triplets through singlet exciton fission[25,31].

We use both optical approaches to show how triplet excitons interact with polariton states, and find that strong coupling creates new radiative channels that are unavailable in the film. The microcavity allows us to extract photons from these 'dark' triplet states by reshaping the potential energy landscape[6–8]. This results in ultra-long-lived polariton emission and the potential for harvesting triplets in devices. The underlying mechanism, based on the widespread phenomenon of excited-state mixing, also opens the way to using strong light-matter coupling to manipulate dipole-forbidden, formally dark states.

**Results**

A common way to study triplet excitons is through delayed fluorescence, which occurs through the spin-allowed conversion of two triplets into a singlet exciton, known as 'triplet-triplet annihilation'[25,32]. One of the best-characterised triplet-triplet annihilation systems is the diphenylanthracene/metal-porphyrin blend used for up-conversion[33,34] shown in Fig 2. We depict the photophysics of this system schematically in Figure 2b: directly exciting the Pt-porphyrin at 532nm initiates efficient intersystem crossing (<100fs)[33], producing triplets that can transfer to diphenylanthracene where triplet-triplet annihilation produces 'up-converted' delayed fluorescence.

In order to understand how triplets behave in microcavities, we need to study delayed fluorescence in the solid state, rather than solution. We therefore prepared films of diphenylanthracene/Pt-porphyrin/polystryene blends with a ratio of 50:1:15. The polystyrene is used to aid mixing between the two active materials and reduce film roughness. Films and microcavites were encapsulated in inert atmosphere to protect against oxygen quenching. The absorption of a control, non-cavity film is shown in Figure 2c. Emission behaviour of the films (Supplementary Figs S1,2) is consistent with literature[33,34]. As expected, excitation within the Pt-porphyrin band (532nm) produces up-converted diphenylanthracene fluorescence (400-500nm). However, as with other solid-state up-conversion systems, we also observe strong Pt-porphyrin phosphorescence (650nm) due to phase separation[33,34].

Figure 2c shows a reflectivity map of a microcavity containing the diphenylanthracene blend, as a function of incident angle and wavelength. The dips in microcavity reflectivity never cross the bare exciton energy (blue dashed). This 'anti-crossing' is a signature of strong light-matter coupling and polariton formation, and the absorbing states are thus split into polariton branches. Transfer-matrix modelling based on measured optical parameters confirms strong coupling in this structure (Fig 2c, lines and circles). Microcavity emission originates from the lower polariton branch (Fig 2c, right), whether we excite the diphenylanthracene directly (355 nm, dashed) or the Pt-porphyrin (532nm, shaded). In the latter case emission is due to up-conversion through triplet-triplet annihilation.

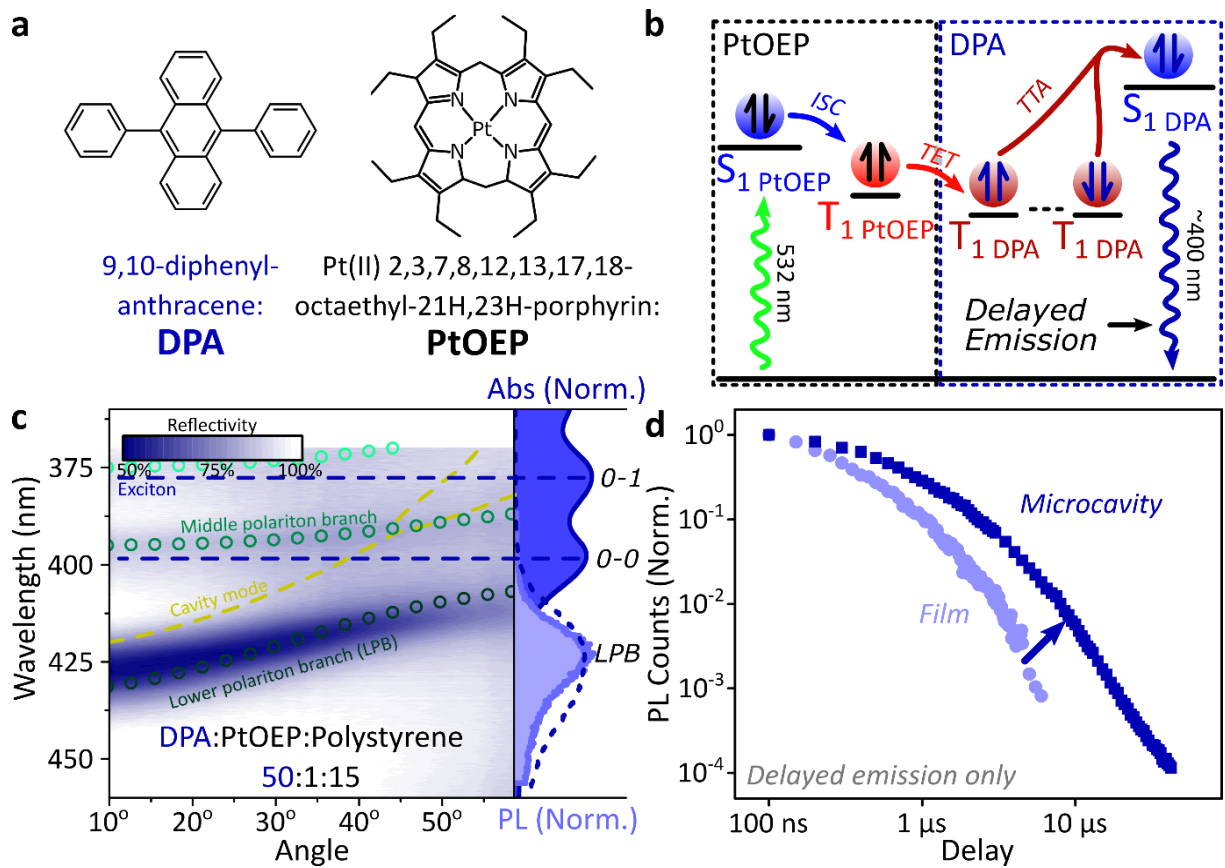

**Fig 2| Sensitised photon up-conversion. a** Molecular structures of active components used in photon up-conversion system. **b** Simplified schematic of photon up-conversion, details in main text. ISC: intersystem crossing, TET: triplet energy transfer, TTA: triplet-triplet annihilation. **c** Reflectivity map of photon up-conversion blend within a Ag-Ag microcavity. Comparison with absorption spectrum (right) and transfer matrix modelling (lines, circles) confirms strong coupling, characterised by anti-crossings at the 0-0 and 0-1 energies (dashed). Details of transfer matrix model in Methods. All emission comes from the lower polariton branch (LPB), whether excitation is resonant with diphenylanthracene (355nm, dashed) or PtOEP (532nm, shaded). Emission is collected with a NA=0.76 lens and thus effectively integrates along the entire dispersion (±45°). **d** Decay kinetics of diphenylanthracene/exciton-polariton emission following excitation of PtOEP at 532nm reveal enhanced lifetime in microcavity (dark) vs bare film (light). All emission on these timescales arises from triplet-triplet annihilation. Incident power (film: 50µW, microcavity: 150µW) was chosen to give similar absorbed power in both samples, details in Methods.

We explore how this triplet harvesting process is affected by strong coupling using time-resolved measurements. To correctly compare the film and cavity kinetics, the incident excitation power was scaled to ensure similar absorbed power in both cases (details in Supplementary Fig S3). The lifetime of emission in the microcavity is distinctly longer than in the film (Fig 2d). The predominant species on these timescales are triplet excitons, so we conclude that additional long-lived triplets are harvested in the microcavity. This change in lifetime is surprising and requires further investigation. We noticed that this ternary blend undergoes laser-induced phase segregation, making detailed studies on this system difficult (Supplementary Fig S3). We therefore apply the same approach to a simpler system with a single active component.

Diketopyrrolopyrrole thiophene (DPPT, Fig 3a) is the base unit for polymers exhibiting high charge-carrier mobility in thin-film transistors, recently used in electrical-injection polariton OLEDs[17]. DPPT monomers are also known to undergo intersystem crossing in the solid state (Fig 3b)[35]. Films were prepared containing DPPT dispersed in polystyrene matrix (1:4 DPPT:polystyrene). Reference photoluminescence measurements on these films reveal delayed fluorescence which is quenched by oxygen (Supplementary Fig S6). This result, together with a non-linear intensity dependence (Supplementary Fig S6), suggests the weak delayed fluorescence in DPPT results from bimolecular triplet-triplet annihilation. All subsequent measurements were performed on films or microcavities encapsulated in an oxygen-free environment.

Within DPPT-based microcavities, Figure 3c, we observe a clear anti-crossing at the 0-0 peak, while the second peak in the absorption appears to be in the weak/intermediate-coupling regime. Similar to the diphenylanthracene cavities, we attribute the anti-crossing states to polariton branches, and emission is again entirely from the lower polariton branch. Comparison of the film and microcavity emission kinetics in Figure 3d reveals that the prompt fluorescence dynamics remain unchanged. However, delayed fluorescence from triplet-triplet annihilation is once again longer-lived in the microcavity. By contrast, in a reference material INDB in which we observe no contribution to emission in the bare film from triplet-triplet annihilation, we also observe no enhancement of long-lived emission in microcavities (Supplementary Figs S7-10). Likewise, when we quench the triplets in DPPT through exposure to oxygen, the enhancement observed in Figure 3d disappears (Supplementary Fig S6).

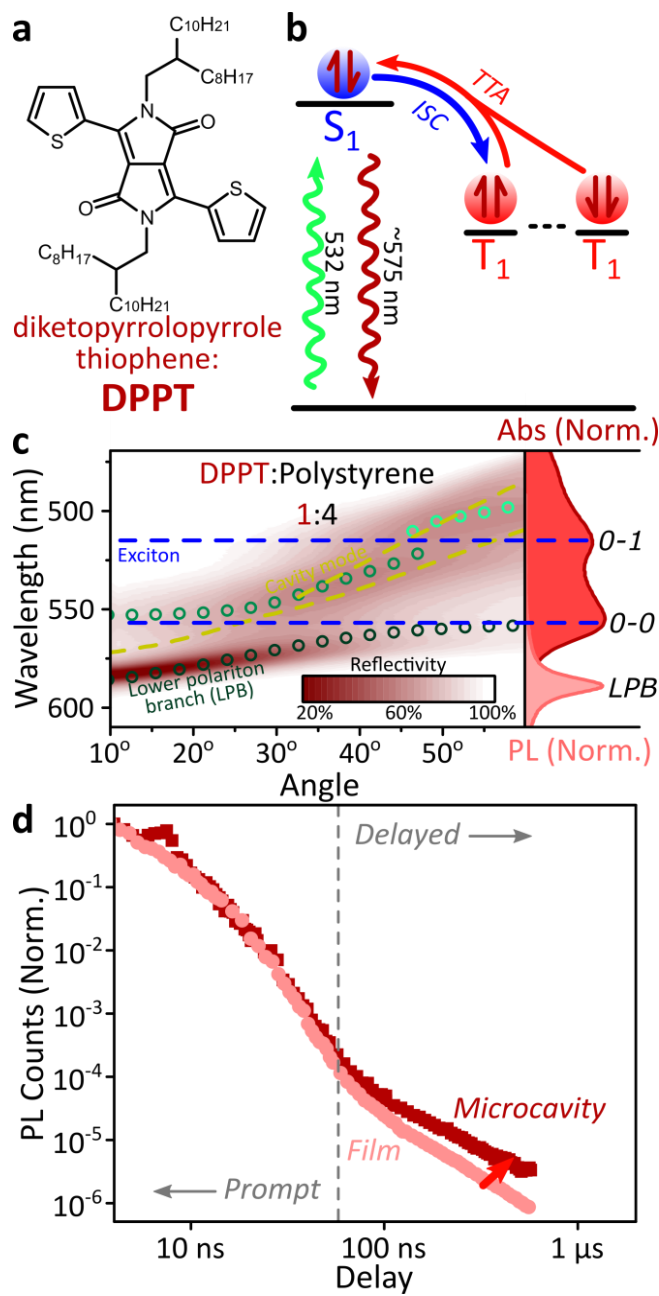

**Fig 3| Delayed emission in single-component film. a** Molecular structure of DPPT. **b** Simplified schematic of DPPT film photophysics, details in main text. ISC: intersystem crossing, TTA: triplet-triplet annihilation. **c** Reflectivity map of DPPT:polystyrene film within a Ag-Ag microcavity. Comparison with absorption spectrum (right) and transfer matrix modelling (lines, circles) confirms strong coupling to the 0-0 transition. All emission arises from the lower polariton branch (LPB). Emission is collected with a NA=0.76 lens and thus effectively integrates along the entire dispersion (±45°). **d** Integrated photoluminescence kinetics over full emission band for bare film (light) and microcavity (dark) following excitation at 532nm. Enhanced microcavity emission matches the 'delayed' regime, in which contributions from triplet-triplet annihilation are significant (Supplementary Fig S6).

To understand the possible origin of this longer-lived emission, we consider the mechanism and spin physics of triplet-triplet annihilation. Emission in this process comes from the encounter of two triplets as they diffuse through the film, forming an 'encounter complex' called a triplet-pair state (TT)[36]. There are nine possible spin combinations of (TT): one spin-0 singlet ($^1$(TT)), three spin-1 triplets ($^3$(TT)) and five spin-2 quintets ($^5$(TT)). Due to spin selection rules, only the singlet-character $^1$(TT) can reform $S_1$ producing delayed fluorescence. Simple spin considerations would imply a maximum quantum efficiency of ~ 11% (1 of 9). In practice, up-conversion efficiencies of >11% have been observed in solution systems, suggested to be due to the higher spin-states such as $^5$(TT) dissociating without significant loss to reform $^1$(TT)[32,37]. We propose that the enhanced long-lived emission in our microcavities can also be explained by considering these high-spin states.

The quintet triplet-pair states have only recently been observed in organic solid-state systems and have been shown to have a lifetime of a few microseconds[38–41]. This long lifetime is presumably because any relaxation to the ground state from $^5$(TT) would be spin-forbidden. On the other hand, $^1$(TT) has a typical lifetime on the order of only 10ns in the absence of fast dissociation into 'free' triplets[42,43]. Moreover, it can directly recombine into $S_1$ in a spin-allowed process. Given this significant difference in pair-state lifetimes, a channel to harvest the quintet ($^5$(TT)) states for delayed fluorescence would significantly increase emission lifetime. To more directly probe this possibility, we study a material system capable of singlet exciton fission, the inverse of triplet-triplet annihilation in which a singlet exciton forms two triplet excitons via $^1$(TT)[25,31,42–44]. The triplet-pair state produced via singlet fission is initially a pure singlet $^1$(TT) which can then evolve into $^5$(TT) on sub-microsecond timescales[38–41]. This well-defined spin evolution, which does not depend on exciton diffusion, allows us to identify the contribution of high-spin states in microcavities.

We make films of TIPS-tetracene, structure in Figure 4a. In polycrystalline TIPS-tetracene films, singlet fission occurs within 50 ps[44], well within our instrument response of 4 ns. The singlet and $^1$(TT) states are very similar in energy resulting in a dynamic equilibrium and delayed fluorescence, Figure 4b. Over time, spin evolution of the bound triplet-pair state forms $^5$(TT)[38] and the films become non-emissive. On very long timescales (>µs), spin dephasing yields independent triplets[38]. We note that even on these long timescales, the predominant triplet-triplet annihilation processes in TIPS-tetracene are geminate[44] because annihilation occurs between triplets formed from the same parent singlet state. As a result, the delayed fluorescence kinetics show no dependence on excitation density.

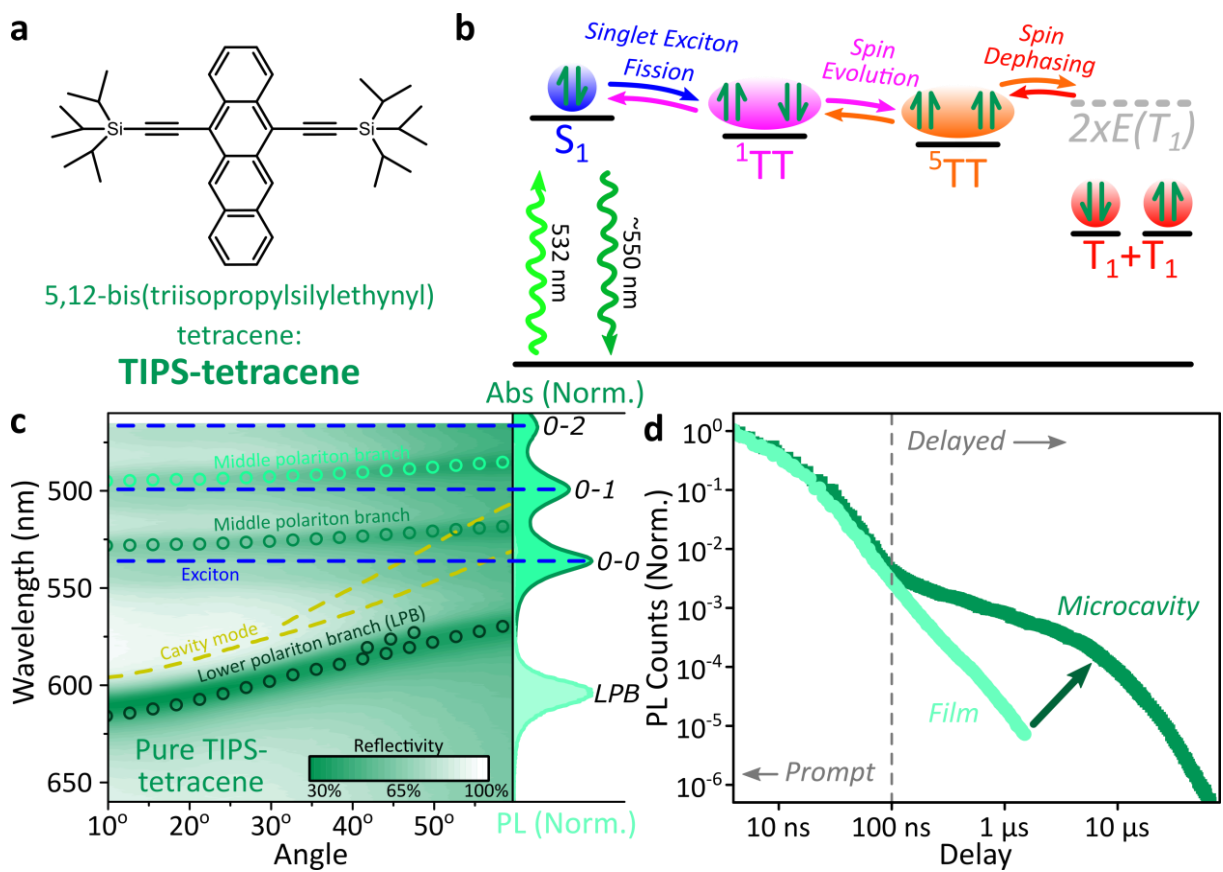

**Fig 4| Singlet fission into bound TT within a microcavity. a** Molecular structure of TIPS-tetracene. **b** Simplified schematic of TIPS-tetracene photophysics, details in main text. All processes are potentially reversible, leading to delayed fluorescence from triplet-triplet annihilation. **c** Reflectivity map of a pure TIPS-tetracene film within a Ag-Ag microcavity. Comparison with absorption spectrum (right) and transfer matrix modelling (lines, circles) confirms strong coupling to multiple transitions. Emission is from the lower polariton branch (LPB). Emission is collected with a NA=0.76 lens and thus effectively integrates along the entire dispersion (±45°). **d** Integrated photoluminescence kinetics over full emission band for bare film (light) and microcavity (dark) following excitation at 532 nm. All emission on these timescales arises from triplet-triplet annihilation.

Within microcavities we observe strong coupling throughout the TIPS-tetracene absorption band and clear polariton branches, Figure 4c. As above, emission is entirely from the lower polariton branch. Relative to the bare film, the prompt microcavity emission is only weakly perturbed, Figure 4d. However, beyond 100ns the microcavity shows a significantly enhanced long-lived tail. As in previous systems, the film and microcavity spectral shapes exhibit negligible evolution over this decay (Supplementary Fig S11). In the film this is because all emission we detect is from $S_1$, populated by triplet-triplet annihilation from $^1$(TT). Likewise, in the microcavity the constant spectral shape indicates that the emission is mediated by the lower polariton state, also populated by triplet-triplet annihilation. However, the clear delineation into two kinetic regimes in Figure 4d suggests that on long times (>100ns) this process follows a distinct pathway unavailable in the film.

To identify this pathway, we compare our results with published data on equivalent TIPS-tetracene polycrystalline films. These are presented in Figure 5. Transient absorption spectroscopy (solid line) has shown that triplet photoinduced absorption signatures generated by singlet fission occur well within our instrument response and do not decay significantly until >10μs[44]. The same sample shows emission with 10ns lifetime, attributed to delayed fluorescence during the $S_1$–$^1$(TT) equilibrium[44]. This decay closely matches our prompt emission decay (open circles). Importantly, the combination of these two data sets shows that it is the character of the triplet-pair states, rather than a loss in population, that causes a decay in delayed fluorescence. This change in character is attributed to spin evolution[38,44]. Indeed, time-resolved electronic paramagnetic resonance spectroscopy shows the presence of quintet states[38], reproduced by the dashed line. Interestingly, the bulk of our microcavity-enhanced emission coincides with this time frame. We conclude that the lower polariton can be directly populated by quintet states. Finally, on still longer timescales, it has been observed that the triplets become independent, with no spin coupling. In this regime, we detect no emission from the microcavity.

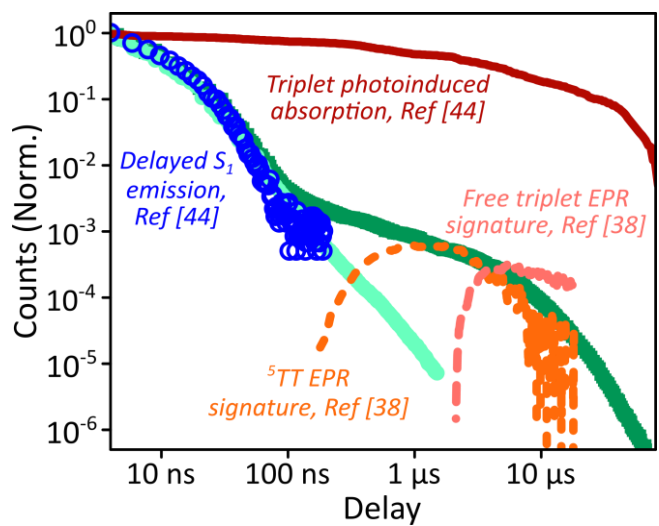

**Fig 5 | Identification of the $^5$TT contribution.** Reproduction of the TIPS-tetracene film and microcavity emission kinetics from Fig 4d, compared with time-correlated single photon counting[44] (circles), transient absorption[44] (solid) and electron paramagnetic resonance[38] (dashed) kinetics previously reported for similar polycrystalline films. The long-time enhancement in microcavities coincides with observations of high-spin $^5$TT.

## Discussion

In all the systems we have studied in which triplets are formed, triplet-triplet annihilation leads to longer-lived emission in the microcavity compared with the bare film. Based on the model system TIPS-tetracene, we suggest that in all cases, the enhanced emission comes from harvesting $^5$(TT) into the lower polariton. Such a result is unexpected as direct coupling between the quintet and the polariton should be spin- and symmetry-forbidden. The mechanism for the enhanced quintet-based emission is therefore unclear. We propose that it can be explained through mixing between states of different character, i.e. the 'adiabatic picture' of organic photophysics.

It is common in work with organic materials to treat systems in terms of their pure or 'diabatic' electronic states, e.g. tightly bound singlet Frenkel exciton $S_1$ or charge-transfer exciton CT. Importantly, this picture is typically applied in the exciton-polariton field, in which the 'exciton' that strongly couples is treated as the diabatic $S_1$ state. However, the states observed in organic systems are not pure diabatic states; they contain admixtures of other states which become significant when the different diabatic states come close in energy. This is a critical distinction from inorganic exciton-polaritons that has not yet been addressed in the literature.

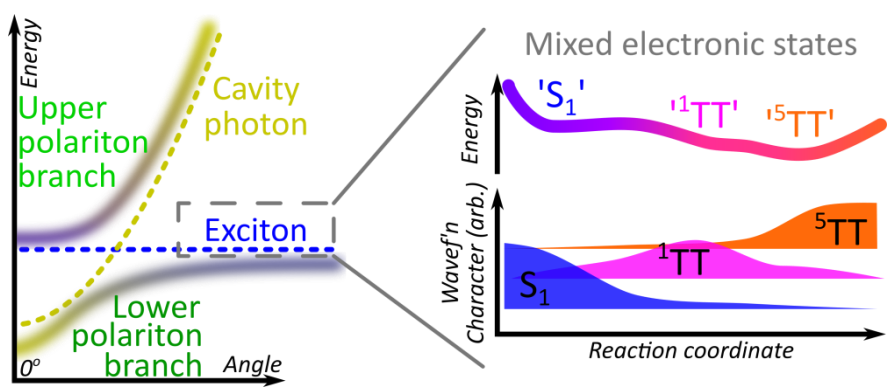

**Fig 6| Mixed electronic states within a microcavity.** Illustrative schematic of the energy landscape and mixed electronic states that contribute to the 'exciton'. The exciton treated in strong coupling is not necessarily a pure diabatic state in organic semiconductor films. The character of the state varies on a continuum along a generalised reaction coordinate. The adiabatic states actually observed (in quotations, top right) are named according to their predominant wavefunction character from the corresponding diabatic states (bottom right), but other configurations also contribute to the total wavefunction. These can become important and must be considered when '$S_1$' is coupled to light to form polaritons.

The existence of this state mixing is a critical driver in photophysical processes such as ultrafast intersystem crossing[45], thermally activated delayed fluorescence[46], singlet exciton fission[42,44] and its reverse, triplet-triplet annihilation. For example, in singlet fission the nominal $S_1$ state contains both CT and TT character, in some cases such as pentacene films to a surprisingly high degree (circa 50% CT)[47]. By the same token, the $^1$(TT) formed by singlet fission carries admixture of $S_1$ and CT, enabling fast singlet fission[42,44], efficient equilibration between $S_1$ and $^1$(TT)[44], direct $^1$(TT) emission[42,43] and triplet-pair binding[42–44,48,49]. Similar mixing likely enables the conversion from triplets to $S_1$ via triplet-triplet annihilation.

The implications of this for strong coupling are profound; when the photon couples to $S_1$ to form polaritons, it is in fact coupling to all the states that mix with $S_1$. Consequently, all states that mix with $S_1$ may acquire some photonic character. In polaritonic systems in which triplet-triplet annihilation occurs, this admixture of photon into the triplet-pair states creates a pathway for scattering into the radiative lower polariton branch and thus a route to harvest light from nominally dark diabatic states. Given the negligible intrinsic emission from these states, we suggest the scattering occurs through a vibrational mechanism[50], which can be highly efficient over the long quintet lifetime. While the above explains any increased microcavity emission from $^1$(TT), it does not on its own explain our observation of $^5$(TT) harvesting. However, it is generally accepted[36] that in the weak spin-interaction (exchange coupling) regime[36,38–41] the spin-singlet and spin-quintet TT states are mixed. Therefore, if $^1$(TT) has some photon component, $^5$(TT) must also acquire a photon component. Hence, in the strong exciton-photon coupling regime, mixed states gain weak but non-zero photon character which creates a channel to populate the emissive lower polariton branch.

We note that the quintet state kinetic we reproduce in Fig 5 corresponds to the signature of strongly spin-interacting (exchange-coupled) quintet states, rather than weakly spin-interacting[38]. We suggest that either the kinetics of weak- and strong-spin-interacting quintets have similar lifetimes or the dynamical process leading to strong spin-interaction is reversible.

Our model does not allow for an emission contribution from states that do not mix, even indirectly, with $S_1$. On the very long time (>10μs) in Fig 5, we see no emission despite an expected population of bound triplets. This is because these spin-independent triplets no longer couple to $S_1$ and therefore do not mix with the photon. By the same token, we detect no polariton emission from isolated triplets in DPPT, diphenylanthracene or Pt-porphyrin as they are not coupled to the photon mode. This shows that the ability to mix with the photon mode is an essential characteristic of squeezing light out of formally dark diabatic states.

Interestingly, this behaviour could allow for an improvement in solid-state up-conversion. In an optimised fluorescence up-conversion system, the ability to directly harvest $^5$(TT) encounter complexes in the weak spin-interaction regime should boost the maximum efficiency. At the same time, the resulting up-converted polariton emission would be well-directed, thereby simplifying collection. A similar mechanism could be used in electrically-injected polariton LEDs and lasers, where triplet excitons constitute 75% of the population and triplet-triplet annihilation could be used to harvest them. The ability of these very long-lived states to populate the lower polariton can also enable new applications in polaritonic physics. For example, it may be possible to use such a reservoir of non-coupled states to feed a polariton condensate, greatly increasing its effective lifetime. This may be equivalent to the continuous pumping of exciton reservoir states in GaAs microcavities to continually repopulate the polariton condensate[10]. Such long-lived condenstates would be important for practical applications of room-temperature polariton condensation. This concept also vastly expands the scope of microcavity-controlled matter, which seeks to alter material properties and light-induced dynamics through strong light-matter coupling[1–8]. The interactions implicit in the adiabatic picture means that strong coupling may perturb not only the state that dominates the absorption spectrum, but also any states that mix with it. It presents new opportunities to control any process in which an absorbing state mixes with others, for example charge transfer, biological light harvesting, energy transfer and intersystem crossing.

**Acknowledgements**

This work was supported by the Engineering and Physical Sciences Research Council, U.K. (Grant Number EP/M025330/1, 'Hybrid Polaritonics'), the University of Sheffield and EU project 679789-455 'CONTREX'.

**Author Contributions**

DWP, HC and AJM performed the experiments. RJ performed transfer-matrix modelling. AL, KJF, JA and HB synthesised materials. DWP, JC and AJM analysed the data and wrote the manuscript. AJM conceived the project and supervised it with DGL and JC. All authors discussed the results and commented on the manuscript.

**Competing Interests**

The authors declare no competing financial interests.

**Supplementary Information**

Supplementary information to the text can be provided on request- please contact corresponding authors.